\documentclass[notitlepage,
 amsmath,amssymb,
 aps,nofootinbib,
]{revtex4-1}

\pdfoutput=1

\usepackage{graphicx}
\usepackage{dcolumn}
\usepackage{bm}

\usepackage{amssymb}
\usepackage{amsmath}
\usepackage{epstopdf}
\usepackage{xcolor}
\usepackage{mathtools}
\usepackage{comment}

\def\hhref#1{\href{http://arxiv.org/abs/#1}{arXiv:#1}} 
\usepackage{hyperref}

\newcommand{\W}{\mathcal{W}}
\newcommand{\brac}[1]{\left( #1 \right)}
\newcommand{\Brac}[1]{\left[ #1 \right]}
\newcommand{\D}{\Delta}
\def\hhref#1{\href{http://arxiv.org/abs/#1}{arXiv:#1}} 
\makeatletter
\renewenvironment{cases}[1][l]{\matrix@check\cases\env@cases{#1}}{\endarray\right.}
\def\env@cases#1{%
  \let\@ifnextchar\new@ifnextchar
  \left\lbrace\def\arraystretch{1.2}%
  \array{@{}#1@{\quad}l@{}}}
\makeatother

\begin{document}


\title{Non-Perturbative Large $N$ Trans-series for the Gross-Witten-Wadia Beta Function}

\author{Anees Ahmed and Gerald V. Dunne}

\affiliation{Department of Physics, University of Connecticut, Storrs, CT 06269, USA}


\begin{abstract}
We describe the non-perturbative trans-series, at both weak- and strong-coupling, of the large $N$ approximation to the beta function of the Gross-Witten-Wadia unitary matrix model. This system models a running coupling, and the structure of the trans-series changes as one crosses the large $N$ phase transition. The perturbative beta function acquires a non-perturbative trans-series completion at large but finite $N$ in the 't Hooft limit, as does the running coupling.
\end{abstract}

\maketitle

\section{Introduction}

One of the big puzzles concerning resurgent asymptotics in QFT \cite{Dunne:2016nmc} is how it applies to the situation where the coupling is not fixed, but runs with the scale. In this short note, we explore this phenomenon in a simple solvable model, the Gross-Witten-Wadia (GWW) unitary matrix model \cite{gw,wadia}, which mimics a running coupling through the dependence on the lattice plaquette scale. The form of the resurgent structure changes as one crosses the large $N$ phase transition. The GWW unitary matrix model is a one-plaquette model of $2d$ Yang-Mills theory, and is defined by the partition function \cite{gw,wadia}:
\begin{eqnarray}
Z(t, N)=\int_{U(N)} DU \, \exp\left[\frac{N}{2\,t} 
{\rm tr}
\left(U+U^\dagger\right)\right]
\label{eq:z}
\end{eqnarray}
Here $t\equiv N g^2/2$ is the 't Hooft coupling. 
The GWW model has a third-order phase transition at infinite $N$, as the specific heat develops a cusp at $t=1$. This large $N$ third order phase transition occurs in many related examples in physics and mathematics \cite{brezin-wadia,Rossi:1996hs,2dgravity,Douglas:1993iia,matytsin,forrester_book,oxford,marcos-book,David:1990ge,Tracy:1993xj,Gross:1992tu}.

For any $N$, the partition function in (\ref{eq:z}) can be compactly expressed as a Toeplitz determinant \cite{Rossi:1996hs}:
\begin{eqnarray}
Z(t, N) =\det \left[ I_{j-k} \left(\frac{N}{t}\right)\right]_{j, k =1, \dots , N}
\label{eq:z1}
\end{eqnarray}
where $I_j$ is the modified Bessel function. While this formula is explicit, the determinant structure makes it of limited use for studying the large $N$ limit. Many alternative techniques have been developed to analyze the large $N$ limit \cite{David:1990ge,brezin-wadia,Rossi:1996hs,2dgravity,forrester_book,oxford,marcos-book}, including the double-scaling limit described by the universal Tracy-Widom form \cite{Tracy:1993xj}. Resurgent asymptotics for the large $N$ limit in matrix models was introduced in \cite{marino-matrix}, using the pre-string difference equation. To study the analytic continuation of the large $N$ trans-series structure, where $N$ becomes complex, one can alternatively map the GWW model to a Painlev\'e III equation (in terms of the 't Hooft coupling $t$), in which $N$ appears as a parameter \cite{ad}. The familiar double-scaling limit of the GWW model arises as the well-known coalescence limit reducing Painlev\'e III to Painlev\'e II \cite{dlmf:ps}. In this paper, we extend this Painlev\'e-based approach to the analysis of the beta function of the GWW model, explaining the form of the large $N$ trans-series, at both weak and strong coupling.

\subsection{Running Coupling and Beta Function}
The running coupling is defined \cite{gw} by reintroducing a length scale (the lattice spacing $a$) into the Wilson loop via the definition
\begin{eqnarray}
\W(t, N) \equiv \exp\left[-a^2\, \Sigma\right]
\label{eq:wilson}
\end{eqnarray}
Keeping the string tension $\Sigma$ fixed therefore defines $t=t(a, N)$ as a function of the scale $a$. This running coupling $t(a, N)$ can be obtained by inversion of the expression \footnote{Note that for any finite $N$, the relation between the 't Hooft coupling $t$ and the lattice scale $a$ is monotonic.}
\begin{eqnarray}
a^2\equiv -\frac{1}{\Sigma}\, \ln \W(t, N)
\label{eq:a}
\end{eqnarray}
The beta function is then defined \cite{gw}:
\begin{eqnarray}
\beta(t, N)=-\frac{\partial\, t(a, N)}{\partial \ln \, a}
\label{eq:beta}
\end{eqnarray}
From now on, we set the string tension $\Sigma=1$, absorbing it into the units of $a$.

\begin{figure}[h]
\centering{\includegraphics[scale=.7]{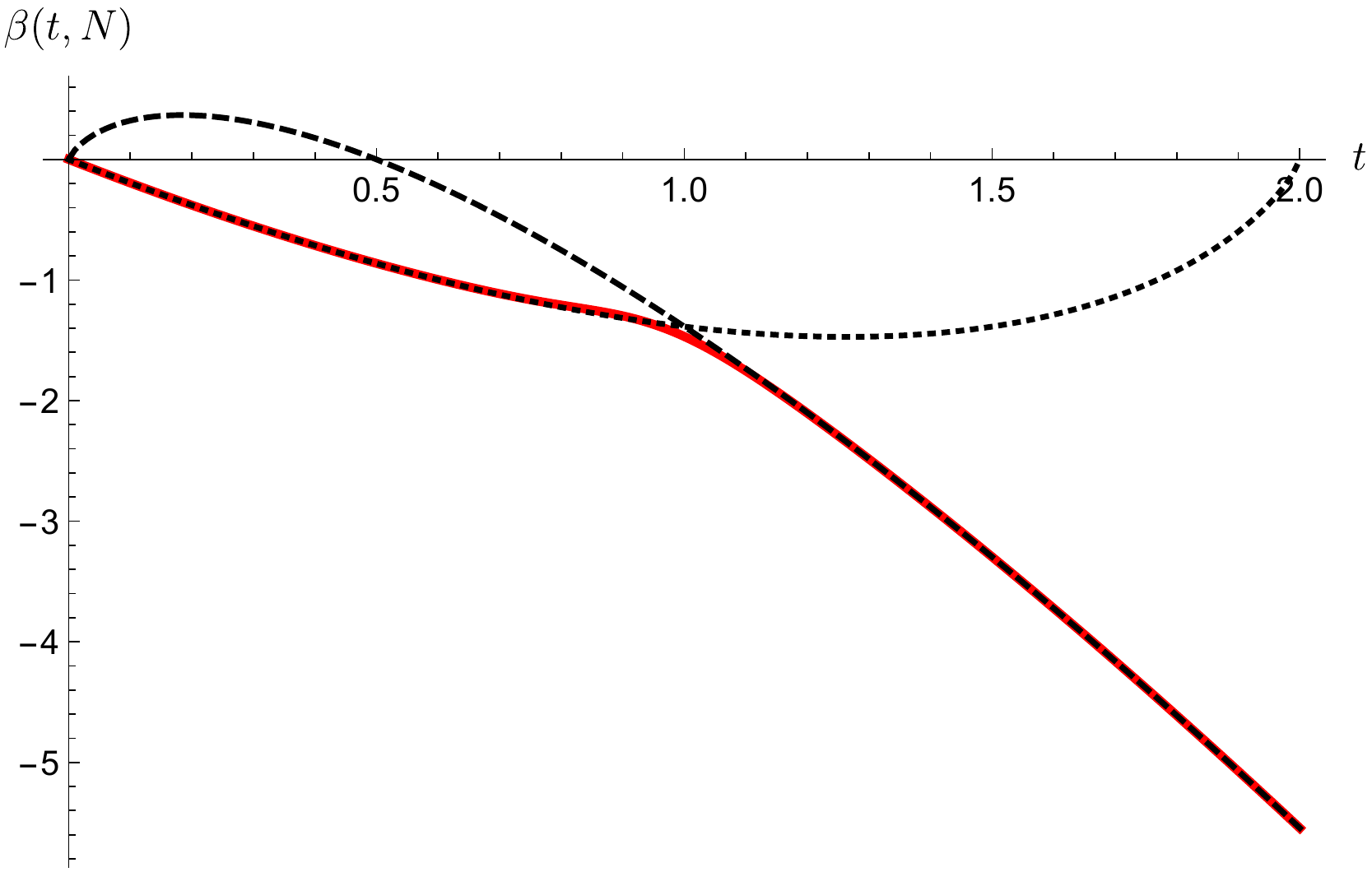}}
\caption{Plot of the GWW beta function $\beta(t, N)$ in (\ref{eq:beta2}). The red solid curve shows the exact beta function for $N=20$. The dashed and dotted lines show the strong-coupling and weak-coupling approximations, respectively, at infinite $N$, from (\ref{eq:betaInfN}). The infinite $N$ approximations show spurious zeros at $t=1/2$ and $t=2$, but in fact the true beta function has a single zero at $t=0$. As $N\to\infty$, the jump at $t=1$ shown in the red curve becomes a cusp, indicating the $N=\infty$ third-order phase transition \cite{gw,wadia}, and the beta function curve jumps from the infinite $N$ strong-coupling form to the infinite $N$ weak-coupling form as $t$ decreases through the phase transition. See Fig. \ref{fig:betaKink} for a close-up of the cusp at $t=1$. The finite $N$ corrections, which produce this jump, are described in Sec. \ref{sec:trans-beta} in the form of a large $N$ trans-series. }
\label{fig:beta}
\end{figure}

\begin{figure}[!t]
    \centering{\includegraphics[scale=1.1]{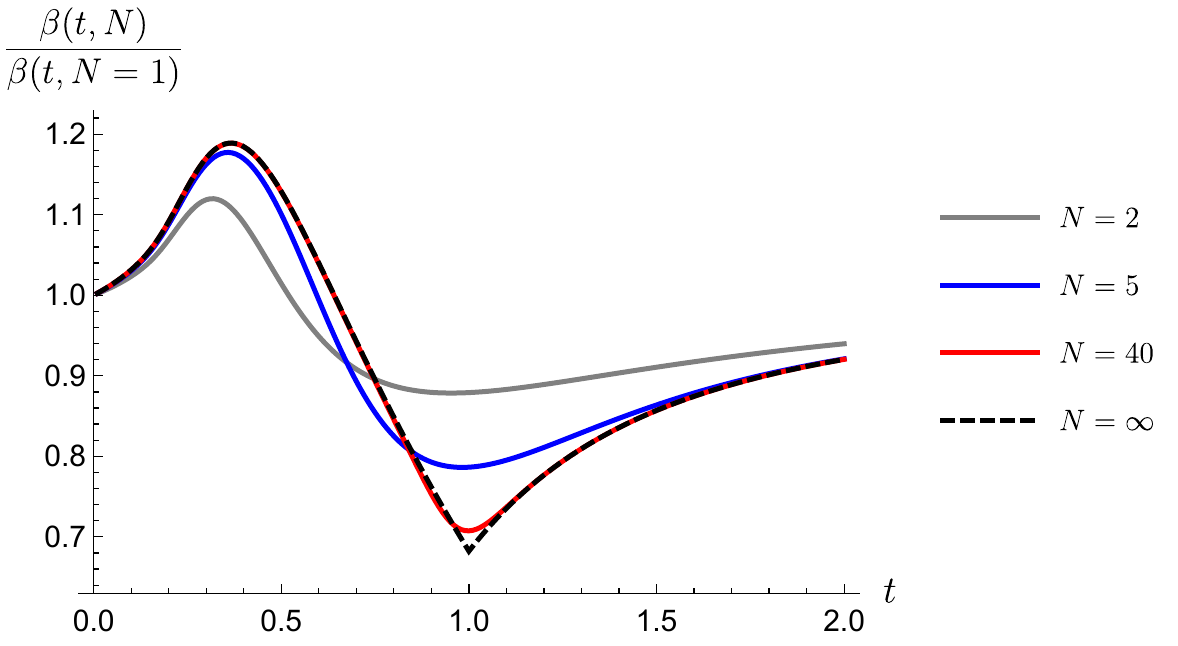}}
    \caption{Plot of the ratio of the beta function $\beta(t, N)$ to the $N=1$ beta function, which shows the development, as $N$ increases, of the kink at the $N=\infty$ phase transition point $t=1$.}
    \label{fig:betaKink}
\end{figure}

\begin{figure}[!t]
    \centering{\includegraphics[scale=.8]{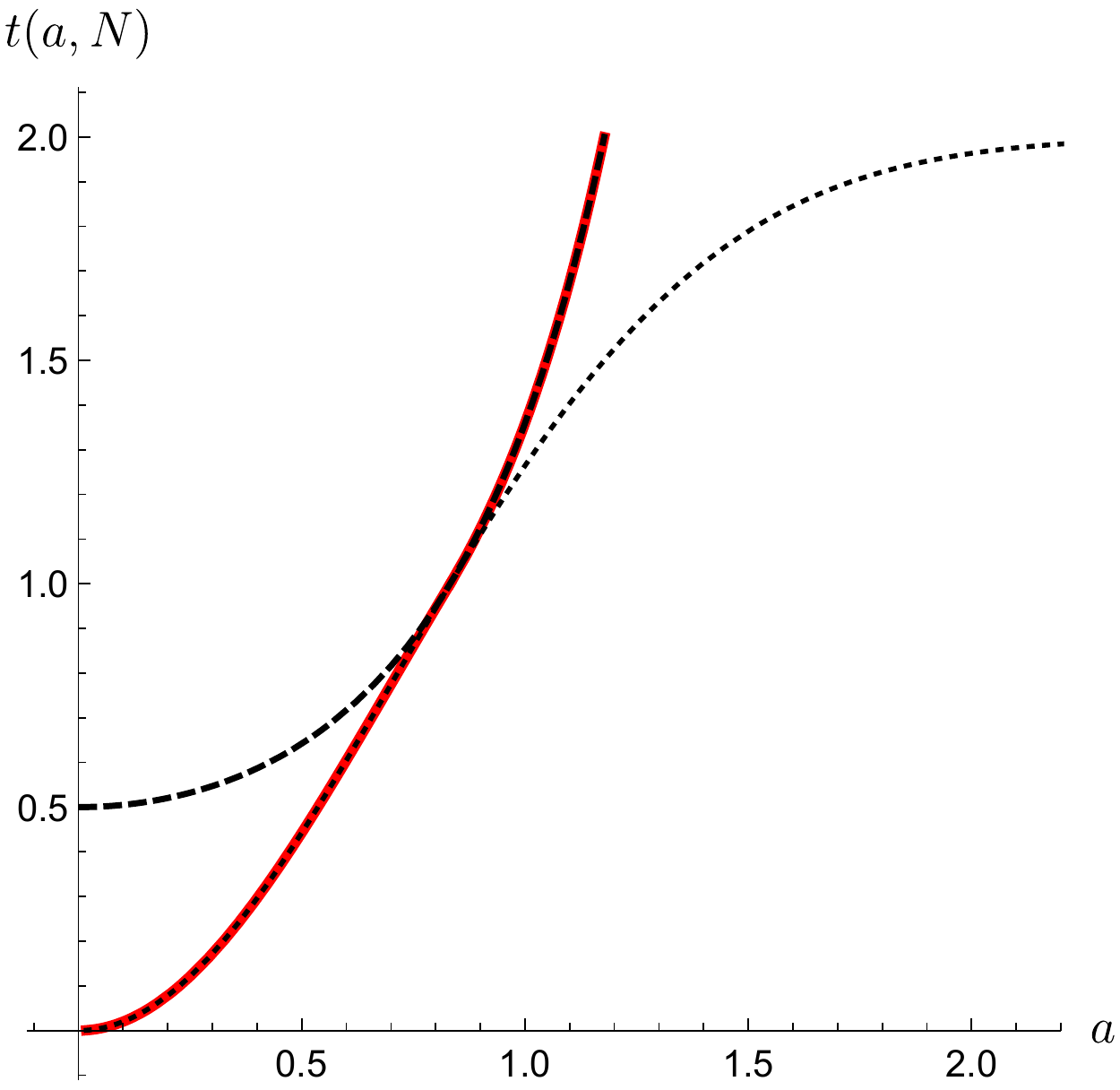}}
    \caption{The running coupling $t(a, N)$ [red solid curve] as a function of the lattice scale $a$, for $N=20$. The dashed and dotted curves show the strong coupling and weak coupling behavior, respectively, at infinite $N$. The true behavior of the running coupling jumps from one asymptotic curve to the other, near the $N=\infty$ phase transition point: $t=1$. The finite $N$ corrections, in the form of a large $N$ trans-series, are described in Section \ref{sec:trans-running}. }
       \label{fig:running}
\end{figure}

At infinite $N$, the Wilson loop at strong and weak coupling is \cite{gw}:
\begin{eqnarray}
\W (t, N) \xrightarrow{N \to \infty} \begin{cases}[r]
\dfrac{1}{2\,t} \text{\qquad\quad strong coupling, } t\geq 1\\
    1-\dfrac{t}{2}  \text{\qquad weak coupling, } t \leq 1
    \end{cases}
\label{eq:w-infinite-n}
\end{eqnarray}
Therefore, at infinite $N$ the running coupling $t(a)$ is:
\begin{eqnarray}
t(a, N) \xrightarrow{N \to \infty} \begin{cases}[r]
    \dfrac{1}{2} \exp\left[ a^2\right] \text{\qquad\qquad strong coupling, } t\geq 1\\
   2\left(1- \exp\left[- a^2\right]\right)  \text{\qquad weak coupling, } t \leq 1
    \end{cases}
\label{eq:t-infinite-n}
\end{eqnarray}
and the beta function is:
\begin{equation} \label{eq:betaInfN}
	\beta(t,N) \xrightarrow{N \to \infty} \begin{cases}[r]
     -2t \ln (2t) \text{\qquad \qquad \quad strong coupling, } t \geq 1 \\
    4 \brac{1- \dfrac{t}{2}} \ln \brac{1- \dfrac{t}{2}}   \text{\quad weak coupling, } t \leq 1
    \end{cases}
\end{equation}
Gross and Witten observed that if one only had the infinite $N$ expressions at either weak or strong coupling, one might erroneously deduce the existence of spurious zeros of the beta function. See Figures \ref{fig:beta} and \ref{fig:betaKink}. Similarly for the running coupling, one might deduce the incorrect behavior at small or large $a$, starting from the other limit at $N=\infty$. See Figure \ref{fig:running}. The resolution of course is that infinite $N$ should be approached from finite $N$, with suitable large $N$ corrections included. In the next Sections we show that these finite $N$ corrections yield non-perturbative trans-series expressions both for the beta function and for the running coupling, and when these are included, the weak coupling expressions match consistently to the strong-coupling expressions. The kink in the beta function, indicating the third order phase transition, develops at $N=\infty$. See Figures \ref{fig:beta} and \ref{fig:betaKink}.

\section{Large N Trans-series for the Beta function}
\label{sec:trans-beta}

From the definition (\ref{eq:a}), for any $N$, we compute $\partial_t a$ and invert, in order to express the beta function in terms of the Wilson loop:
\begin{equation}	
\beta(t, N)=-2\frac{\W(t, N)\, \ln\, \W(t, N)}{\partial_t\, \W(t, N)} =\frac{-2}{\partial_t \left(\ln\, \ln \W(t, N)\right)}
\label{eq:beta2}
\end{equation}
This implies that the beta function $\beta(t, N)$ inherits its non-perturbative trans-series structure directly from the trans-series structure of the Wilson loop $\W(t, N)$. The large $N$ trans-series for $\W(t, N)$ was studied in \cite{ad}, showing how the form of the trans-series changes across the phase transition at $t=1$. Related changes therefore occur for the beta function. For other discussions of non-perturbative effects for the GWW Wilson loop, see \cite{Okuyama:2017pil,Alfinito:2017hsh}.

We briefly review some relevant results from \cite{ad}. The non-perturbative trans-series form of $\W(t, N)$ at any $N$ is efficiently expressed in terms of a solution to a Painlev\'e III equation. Define $\Delta(t, N)$ as the expectation value of the determinant in the Gross-Witten-Wadia model:
\begin{eqnarray}
\Delta(t, N)\equiv \langle \det U\rangle 
\label{eq:delta}
\end{eqnarray}
Then $\W(t, N)$ is related to $\Delta(t, N)$, for any $N$, as:
\begin{equation}
	\W(t, N) = \frac{1}{2t} \Brac{1 - \D^2 - \frac{t^2}{1-\D^2} \brac{\frac{t^2 (\partial_t \D)^2}{N^2} -1}} - \frac{t}{2}
    \label{eq:delta-w}
\end{equation}
The expectation value $\Delta(t, N)$ satisfies the following nonlinear ordinary differential equation, as a function of the 't Hooft coupling $t$, for any value of $N$ \cite{Rossi:1982vw,Rossi:1996hs,ad}:
\begin{equation}
	t^2\partial_t^2 \D + \frac{N^2 \Delta}{t^2}\left(1- \Delta^2 \right)  = \frac{\Delta}{1-\Delta^2}\left(N^2 -t^2 \left(\partial_t \Delta\right)^2\right)
    \label{eq:rossi}
\end{equation}
Notice that $N$ appears as a parameter in this equation, thereby enabling a simple analysis of the large $N$ limit, including analytic continuation in $N$.
The equation (\ref{eq:rossi}) is directly related to the Painlev\'e III equation, and standard resurgent asymptotic techniques \cite{costin-book} permit the development of explicit trans-series expansions in various limits: for example, weak or strong 't Hooft coupling \cite{ad}.

Combining (\ref{eq:beta2}) and (\ref{eq:delta-w}), the GWW beta function can also be expressed in terms of $\Delta(t, N)$:
\begin{eqnarray} 
\beta(t,N) &=& -2 t \left(1-\frac{2 \left(\Delta ^2-1\right)^2}{\Delta ^4-\left(t^2+2\right) \Delta ^2+1+ t^4 (\partial_t \D)^2/N^2}\right) \nonumber\\
&&\times\ln \Brac{ \frac{1}{2t} \brac{1-\Delta ^2- \frac{t^2 }{1-\Delta ^2} \left(\frac{ t^2 (\partial_t\D)^2}{N^2}-1\right)} - \frac{t}{2} }
\label{eq:betaFromDelta}
\end{eqnarray}
For example, from (\ref{eq:rossi}) we see that at infinite $N$
\begin{equation} \label{eq:deltaInfN}
	\Delta(t,N) \xrightarrow{N \to \infty} \begin{cases}[r]
    0 \text{\qquad\quad strong coupling, } t \geq 1\\
     \sqrt{1-t}  \text{\qquad\quad weak coupling, } t \leq 1
    \end{cases}
\end{equation}
from which follows the infinite $N$ beta function in (\ref{eq:betaInfN}).

The correspondence (\ref{eq:betaFromDelta}) means that we can use the trans-series structure of $\Delta(t, N)$ to study the trans-series structure of $\beta(t, N)$. And since the trans-series expansions of $\Delta(t, N)$ were shown in \cite{ad} to display concrete resurgence relations between different non-perturbative sectors in the trans-series, it follows that the same is true for the beta function $\beta(t, N)$.

We can also use the relation (\ref{eq:betaFromDelta}) to plot the beta function as a function of coupling, for various values of $N$: see Figures \ref{fig:beta} and \ref{fig:betaKink}. These figures illustrate the fact that for any given $N$, the weak coupling dependence merges consistently with the strong coupling dependence, with a cusp developing at the critical 't Hooft coupling only at $N=\infty$. In particular, it is clear that the zeros of the infinite $N$ beta function at $t=1/2$ and $t=2$ (see Fig. \ref{fig:beta}) are indeed spurious. 

It is instructive to study the leading trans-series corrections to the infinite $N$ beta functions in (\ref{eq:betaInfN}). The form of the trans-series changes across the phase transition, so we illustrate this change of structure by considering the leading contributions at large but finite $N$. Express the Wilson loop for any finite $N$ as
\begin{eqnarray}
\W=\W_{\rm pert}+ \W_{\rm non-pert}
\label{eq:wilson-pnp}
\end{eqnarray}
Keeping the leading power of the non-perturbative term, we obtain the following expression for the beta function:
\begin{eqnarray}
\beta(t, N) &=& -\frac{2 \ln \W_{\rm pert}(t, N)}{\partial_t \ln \W_{\rm pert}(t, N)} + \dfrac{2}{\partial_t\W_\text{pert}} \brac{ \frac{\ln \W_{\rm pert}}{\partial_t \ln \W_{\rm pert}} \partial_t \W_{\rm non-pert} - \ln \left(e\, \W_{\rm pert}\right)  \W_{\rm non-pert} } + \dots
 \nonumber \\
 & \equiv &\beta_{\rm pert}(t, N) + \beta_{\rm non-pert}(t, N)  
\label{eq:beta-trans}
\end{eqnarray}
where the dots refer to higher powers of $\W_{\rm non-pert}$. 

\subsection{Large N expansion at strong 't Hooft coupling}
\label{sec:large-n-strong}

In the strong coupling limit, $\Delta_{\rm pert}$ is identically zero, so $\Delta(t, N)$ is purely non-perturbative \cite{ad}. Consequently, from (\ref{eq:delta-w}) we deduce that the Wilson loop $\W(t, N)$ has only one perturbative term, $\W_{\rm pert}=\frac{1}{2t}$, which is independent of $N$, and equal to the familiar infinite $N$ Wilson loop in (\ref{eq:betaInfN}). At finite $N$, the further corrections are all non-perturbative. Keeping the leading such non-perturbative correction \cite{ad,Okuyama:2017pil,Alfinito:2017hsh},
\begin{equation} \label{eq:w-large-n-strong}
	\resizebox{.98\hsize}{!}{$\W_\text{strong}(t, N) = \dfrac{1}{2 t}  -  \dfrac{t \, e^{-2 N S_\text{strong}(t)}}{ 4 \pi  N^2 (t^2 - 1)}  \left(1-\dfrac{t \left(3+14 t^2\right)}{12 N \left(t^2-1\right)^{3/2}}+\dfrac{t^2 \left(81+804 t^2+340 t^4\right)}{288 N^2 \left(t^2-1\right)^3} + \dots\right)+\dots$}
\end{equation}
where the large $N$ instanton action at strong coupling is
\begin{eqnarray}
S_{\rm strong}(t)=  {\rm arccosh(t)} - \sqrt{1-1/t^2}
\label{eq:s-strong}
\end{eqnarray}
This translates into a non-perturbative large $N$ instanton correction to the infinite $N$ beta function in (\ref{eq:betaInfN}):
\begin{equation} \label{eq:beta-strong}
\begin{aligned}
\beta_\text{strong}(t, N) = -2t \ln (2t) & \\
- \dfrac{1}{N \pi} \frac{2t^2 \ln (2t)}{\sqrt{t^2-1}} & e^{-2 N S_\text{strong}(t)}   \brac{1+\frac{t \left(6 t^2-6-\left(14 t^2-9\right) \ln (2 t)\right)}{12 N \left(t^2-1\right)^{3/2} \ln (2 t)} + \dots} + \dots
\end{aligned}
\end{equation}
Note the appearance of further terms involving $\ln(t)$ in the fluctuations about the leading large $N$ instanton term, consistent with general trans-series structure \cite{costin-book,sauzin,Aniceto:2018bis}.

At any finite $N$, the expression (\ref{eq:beta-strong}) has an unphysical divergence at $t=1$, arising from use of the Debye expansion for the Bessel functions \cite{dlmf:debye}. In \cite{ad}, the leading large $N$ correction for the Wilson loop at strong coupling was calculated more precisely to be:
\begin{eqnarray}
\W_\text{strong}(t, N) \approx \frac{1}{2 t}-\frac{1}{2 t}\left[\left(J_N\left(N/t\right)\right)^2-J_{N-1}\left(N/t\right) J_{N+1}\left(N/t\right)\right]+\dots
\label{eq:w-strong-bessel}
\end{eqnarray}
This leading correction, in terms of Bessel J functions, is exponentially small at large $N$, and represents a resummation of all fluctuations about the leading large $N$ instanton exponential factor in (\ref{eq:w-large-n-strong}). At finite $N$, expression (\ref{eq:w-strong-bessel}) is therefore much more accurate than the conventional large $N$ expression (\ref{eq:w-large-n-strong}) in the vicinity of the large $N$ phase transition, at $t=1$, where instantons and their fluctuations condense \cite{ad,Neuberger:1980as}.

A {\it uniform} large $N$ instanton expression is obtained by using the uniform large $N$ approximation \cite{dlmf-uniform} for the Bessel functions appearing in (\ref{eq:w-strong-bessel}). This is a nonlinear analogue of the uniform WKB approximation, smooth through the transition point for any finite $N$, and expressed in terms of an Airy function rather than an exponential \cite{ad,dlmf-uniform}. Physically, this uniform large $N$ approximation arises from  the merging of two saddles at the large $N$ phase transition. A similar expression, along with a corresponding uniform approximation, can be deduced for the beta function at large $N$, in the strong coupling regime:
\begin{equation} \label{eq:beta-strong-bessel}
\beta_\text{strong}(t, N) \approx -2t \log (2 t) -2t  \left(\left(J_N\left(N/t\right)\right)^2+(2 \log (2t)-1) J_{N-1}\left(N/t\right) J_{N+1}\left(N/t\right)\right)+\dots
\end{equation}

\subsection{Large N expansion at weak 't Hooft coupling} \label{sec:large-n-weak}
In the weak coupling regime, the infinite $N$ expression in (\ref{eq:deltaInfN}), $\Delta\sim\sqrt{1-t}$, receives both perturbative and non-perturbative  corrections at finite $N$: 
\begin{eqnarray}
\Delta(t, N) = \Delta_{\rm pert}(t, N) +  \Delta_{\rm nonpert}(t, N)
\label{eq:delta-weak}
\end{eqnarray}
This structure flows through to the Wilson loop and to the beta function.
\begin{equation}  \label{eq:w-weak}
\begin{aligned}
	\W_\text{weak}(t, N) = &\brac{1-\frac{t}{2} - \frac{t^2}{8 N^2 (1-t)}  + \dots} \\
   &-\dfrac{i}{2 \sqrt{2 \pi } N^{3/2}} \frac{t}{ (1-t)^{1/4}}  e^{-N S_\text{weak}(t)} \brac{1 + \frac{8+12 t+9 t^2}{96 N (1-t)^{3/2}} + \dots}
\end{aligned}
\end{equation}
where the large $N$ instanton action at weak coupling is
\begin{equation}
	S_\text{weak}(t) = \frac{2 \sqrt{1-t}}{t}-2\, {\rm arctanh}\left(\sqrt{1-t}\right)
	\label{eq:s-weak}
\end{equation}
The corresponding large $N$ trans-series expansion for the beta function has the form 
\begin{eqnarray}
	\beta_\text{weak}(t, N) &=& 2 (2-t) \log \left(1-\frac{t}{2}\right) \Bigg[\left(1  - \frac{t \left(t-t^2 + (4-3 t) \ln \left(1-\frac{t}{2}\right)\right)}{4 N^2 (2-t) (1-t)^2 \log \left(1-\frac{t}{2}\right)} + \dots \right)   \nonumber\\
 &&   - i \sqrt{\frac{2}{\pi  N}} \frac{ (1-t)^{1/4}}{t}   e^{-N S_\text{weak}(t)} \left(1 + \dots\right) + \dots \Bigg]
\end{eqnarray}

\subsection{Large N Double-scaling Limit}

It is well known that the double-scaling limit is described by the Painlev\'e II equation \cite{gw,wadia,marino-matrix}. In our approach this can be seen as follows.
In the double-scaling limit, zoomed in to the immediate vicinity of the GWW phase transition at $t=1$, the Rossi equation (\ref{eq:rossi}) reduces to a Painlev\'e II equation in terms of the scaled variable $\kappa$ which measures the scaled deviation from $t=1$:
\begin{eqnarray}
t=1+\frac{\kappa}{N^{2/3}}\qquad,\qquad \Delta(t, N)=\frac{t^{1/3}}{N^{2/3}}\, V(\kappa)
\label{eq:ds}
\end{eqnarray}
Here $V(\kappa)$ is the real Hastings-McLeod solution of the Painlev\'e II equation \cite{marino-matrix,ad}. In this double-scaling limit, the Wilson loop behaves as  
\begin{eqnarray}
\W_\text{double-scaling}(\kappa) \approx \frac{1}{2}-\frac{\kappa}{2\, N^{2/3}} +\frac{\brac{\kappa+V^2(\kappa)}^2-(V^\prime(\kappa))^2}{2\, N^{4/3}}+ \mathcal{O}\left(\frac{1}{N^2}\right)
\label{eq:ds-v}
\end{eqnarray}
and the beta function as 
\begin{eqnarray} 
\beta_\text{double-scaling}(\kappa) &\approx &   -2 \ln (2) \nonumber\\
&& - \frac{4 \ln 2}{N^{2/3}} \bigg(\kappa  \frac{\ln (2 e)}{2 \ln 2} +  V^2(\kappa)   + 2\kappa  V(\kappa) V'(\kappa) + 2 V^3(\kappa)  V'(\kappa)- V'(\kappa) V''(\kappa) \bigg) \nonumber\\
&&
  +\,  \mathcal{O}\brac{\frac{1}{N^{4/3}}}
 \label{eq:ds-beta}
\end{eqnarray}
This matches smoothly to the strong- and weak-coupling sides of the phase transition, as shown for the double-scaling limit of  $\Delta(t, N)$ in \cite{ad}.

\section{Large N Trans-series for the Running Coupling}
\label{sec:trans-running}

At infinite $N$, the running coupling has the form in (\ref{eq:t-infinite-n}). The finite $N$ corrections, described in the previous section for the beta function, lead also to trans-series structures for $t(a, N)$. At strong coupling, where the scale $a$ is large, the corrections are naturally expressed in terms of the Wilson loop, $\W= \exp[-a^2]$; while at weak coupling, where the scale $a$ is small, the corrections are naturally expressed in terms of $1-\W=1- \exp[-a^2]$. The infinite $N$ phase transition occurs at $\W=1/2$. At any finite $N$, the running coupling, $t(a, N)$  solves the scaling equation
\begin{eqnarray}
a\frac{\partial t}{\partial a} =2\frac{\W(t, N)\, \ln \W(t, N)}{\partial_t \W(t, N)}
\label{eq:scale}
\end{eqnarray}
which is both non-linear and non-perturbative.
It is convenient to consider the coupling as a function of the Wilson loop $\W$. At $N = \infty$ we have:
\begin{equation} \label{eq:tFromWInfN}
	t(\W,N) \xrightarrow{N \to \infty} \begin{cases}[r] 
    \dfrac{1}{2\W} \text{\qquad\qquad strong coupling, } \W \leq \frac{1}{2}\\
    2(1- \W)  \text{\qquad weak coupling, } \W \geq \frac{1}{2}
    \end{cases}
\end{equation}
By matching the expansions of $\W$, we deduce the following
large $N$ trans-series structures for $t$ as a function of $\W$ (and hence of $a$)
\begin{equation} \label{eq:tTransInfN}
	t(\W,N) = \begin{cases}[r]
     \dfrac{1}{2 \W} + \sum_{k=1}^\infty e^{-k N \hat{S}_\text{strong}(\W)} \mathcal{F}^{(k)}_\text{strong}(\W,N), \text{\quad strong coupling, } \W \leq \frac{1}{2} \\
    \sum_{k=0}^\infty e^{-k N \hat{S}_\text{weak}(\W)} \mathcal{F}^{(k)}_\text{weak}(\W,N), \text{\qquad\qquad weak coupling, } \W \geq \frac{1}{2}
\end{cases}
\end{equation}
The actions $\hat{S}_\text{strong}(\W)$ and $\hat{S}_\text{weak}(\W)$ are the strong and weak coupling actions $S_\text{strong}(t)$ and $S_\text{weak}(t)$, evaluated at the infinite $N$ values of $t$ as given in \eqref{eq:tFromWInfN}:
\begin{equation}
\begin{aligned}
\hat{S}_\text{strong}(\W) &= S_\text{strong}\brac{\dfrac{1}{2\W}} = {\rm arccosh} \brac{\dfrac{1}{2\W}} - \sqrt{1-4\W^2}  \\
\hat{S}_\text{weak}(\W) &= S_\text{weak}\big(2(1- \W) \big)  = \dfrac{\sqrt{2\W-1}}{1-\W} - 2\, {\rm arctanh} \brac{\sqrt{2\W-1}} 
\end{aligned}
\end{equation}
The leading terms in the strong coupling trans-series \eqref{eq:tTransInfN} read:
\begin{equation}
\begin{aligned}
	t_\text{strong}&(\W,N) \approx  \frac{1}{2\W}  \\
    &- \frac{e^{-N \hat{S}_\text{strong}(\W)} }{4 \pi  N^2 \W \left(1 - 4 \W^2\right)} \brac{1 -\frac{6 \W^2+7}{6 N \left(1-4 \W^2\right)^{3/2}} -\frac{324 \W^4+804 \W^2+85}{72 N^2 \left(1 - 4 \W^2\right)^3} + \dots} + \dots
\end{aligned}
\end{equation}
understood as being expanded in $\W= \exp[-a^2]$. At weak coupling
\begin{equation}
\begin{aligned}
	t_\text{weak}(\W,N) \approx 2 (1-\W) \brac{1 -\frac{1-\W}{2 N^2 (2 \W-1)}  + \frac{\left(8 \W^2+5 \W-9\right) (1-\W)^2}{8 N^4 (2 \W-1)^4} + \dots }  \\
    - \frac{2 i}{\pi N^{3/2}} \frac{1-\W}{\brac{2\W-1}^{1/4}} e^{- N \hat{S}_\text{weak}(\W)} \brac{1 - \frac{9 \W^2+5}{24 N (2 \W-1)^{3/2}} + \dots} + \dots
\end{aligned}
\end{equation}
understood as being expanded in $(1-\W)=(1- \exp[-a^2])$. 

\section{Conclusions}

The Gross-Witten-Wadia unitary matrix model is a one-plaquette model of 2 dimensional lattice Yang-Mills theory, which has the interesting feature of a third-order phase transition at infinite $N$, in addition to a running coupling \cite{gw,wadia}. The perturbative beta function for this model acquires a non-perturbative trans-series completion at large but finite $N$ in the 't Hooft limit, as does the running coupling. The 't Hooft coupling runs with the scale $a$, and the trans-series rearranges itself across the phase transition. Physically, this transition is identified with the condensation of instantons \cite{Neuberger:1980as}, with different kinds of instantons dominating at weak- and strong-coupling \cite{marino-matrix,Buividovich:2015oju,Alvarez:2016rmo}. Technically, the beta function $\beta(t, N)$ can be expressed explicitly in terms of the expectation value $\Delta(t, N)\equiv \langle \det U\rangle$, whose resurgent trans-series structure was studied in detail in \cite{ad}. The beta function $\beta(t, N)$ inherits its trans-series structure from that of $\Delta(t, N)$, and therefore the beta function trans-series also has full resurgent properties, including concrete relations between different instanton sectors. It would be interesting to study further this trans-series structure directly in the renormalization group approach to matrix models \cite{Carlson:1984nb,Brezin:1992yc,Damgaard:1993df,Higuchi:1993nq}.

\bigskip

{\noindent\bf Acknowledgments:}
This material is based upon work supported by the U.S. Department of Energy, Office of Science, Office of High Energy Physics under Award Number DE-SC0010339. 

\end{document}